\documentclass[prd,twocolumn,nofootinbib]{revtex4-1}
\usepackage{graphicx}
\usepackage{bm}
\usepackage[colorlinks=true]{hyperref}
\usepackage{float}
\usepackage{amsmath}
\usepackage{amssymb}
\usepackage{pifont}
\usepackage{gensymb}
\usepackage{multirow}
\usepackage[dvipsnames]{xcolor}
\usepackage[utf8]{inputenc}
\usepackage[normalem]{ulem}
\newcommand\be{\begin{equation}}
\newcommand\ba{\begin{eqnarray}}
\newcommand\ee{\end{equation}}
\newcommand\ea{\end{eqnarray}}
\newcommand\bw{\begin{widetext}}
\newcommand\ew{\end{widetext}}

\newcommand{\order}[1]{\mathcal{O}(#1)}

\makeatletter
\newcommand\footnoteref[1]{\protected@xdef\@thefnmark{\ref{#1}}\@footnotemark}
\makeatother

\definecolor{KellyGreen}{RGB}{76,187,23}

\begin{document}
\title{Parameter Estimation for Tests of General Relativity with the Astrophysical Stochastic Gravitational Wave Background}
\author{Alexander Saffer}
\affiliation{Department of Physics, University of Virginia, Charlottesville, Virginia 22904, USA}
\author{Kent Yagi}
\affiliation{Department of Physics, University of Virginia, Charlottesville, Virginia 22904, USA}

\date{\today}
\begin{abstract}
Recent observations of gravitational waves from binary black holes and neutron stars allow us to probe the strong and dynamical field regime of gravity. On the other hand, a collective signal from many individual, unresolved sources results in what is known as a stochastic background. We here consider probing gravity with such a background from stellar-mass binary black hole mergers. We adopt a simple power-law spectrum and carry out a parameter estimation study with a network of current and future ground-based detectors by including both general relativistic and beyond general relativistic variables. For a network of second-generation detectors, we find that one can place meaningful bounds on the deviation parameter in the gravitational-wave amplitude if it enters at a sufficiently negative post-Newtonian order. However, such future bounds from a stochastic background are weaker than existing bounds from individual sources, such as GW150914 and GW151226. We also find that systematic errors due to mismodeling of the spectrum is much smaller than statistical errors, which justifies our use of the power-law model. Regarding a network of third-generation detectors, we find that the bounds on the deviation parameter from statistical errors improve upon the second-generation case, though systematic errors now dominate the error budget and thus one needs to use a more realistic spectrum model. We conclude that individual sources seem to be more powerful in probing general relativity than the astrophysical stochastic background. 

\end{abstract}
\maketitle

\section{Introduction}
\label{sec:Introduction}
The detection of gravitational waves from individual compact binary sources over the past few years have added evidence to that existence of gravitational waves (GWs) in accordance with Einstein's theory of general relativity (GR)~\cite{Abbott:2016blz,TheLIGOScientific:2016pea,Abbott:2016nmj,GBM:2017lvd,TheLIGOScientific:2017qsa,LIGOScientific:2018mvr}. 
These GWs allow for tests of GR as they probe the spacetime near compact objects, otherwise known as the strong and dynamical gravity regime~\cite{Yunes:2013dva,Berti:2015itd,TheLIGOScientific:2016src,Yunes:2016jcc,Abbott:2018lct,LIGOScientific:2019fpa}.  Prior to this, solar system tests~\cite{Will:2014kxa}, table-top experiments~\cite{Adelberger:2006dh,Adelberger:2009zz}, radio pulsar observations~\cite{Stairs:2003eg,Wex:2014nva} and cosmological observations~\cite{Jain:2010ka,Clifton:2011jh,Koyama:2015vza} had been used to place constraints on modified gravity theories.  However, these tests probe the weak or static field regime and showed little evidence for non-GR effects. For example, even the most relativistic binary system has an orbital velocity of $v/c\sim2\times 10^{-3}$~\cite{Burgay:2003jj}.  The detection of GWs from a coalescing binary system was the first step towards probing gravity in a highly relativistic region.

Following the detection of the first GW signal GW150914, several tests were performed and showed agreement between the GR and the detected signal~\cite{TheLIGOScientific:2016src,Yunes:2016jcc,Abbott:2018lct,LIGOScientific:2019fpa}, thus placing stronger bounds on several modified theories of gravity.  
One useful formalism for testing GR with GWs in a theory-agnostic way is the parameterized post-Einsteinian (ppE) formalism~\cite{Yunes:2009ke}.
Rather than working in a specific modified theory of gravity, the ppE formalism allows for deviations in the waveform that arise from those in the binding energy of a binary, Kepler's Law, and GW luminosity. Generally, the ppE formalism can account for deviations in both the amplitude and phase corrections of GWs. Known mapping exists between ppE and non-GR parameters in a given theory~\cite{Tahura:2018zuq}. For corrections entering in the inspiral part of the waveform, there is a one-to-one correspondence between ppE and generalized IMRPhenom formalism~\cite{TheLIGOScientific:2016src,Abbott:2018lct,LIGOScientific:2019fpa} used by the LIGO/Virgo Collaborations (LVC)~\cite{Yunes:2016jcc}. 

While much attention has been given to individual events, there exist binary signals whose signal-to-noise ratios (SNRs) are too small to be detected, and make up a stochastic GW background (sGWB)~\cite{Allen:1997ad,Romano:2016dpx}\footnote{For an overview of sources of the sGWB, see~\cite{Kuroyanagi:2018csn}.  One example of what could be gleamed from an understanding of the sGWB is the history of the early universe, since GW can propagate through spacetime without loss of information~\cite{Starobinsky:1979ty,Maggiore:1999vm,Easther:2006gt,Khlebnikov:1997di,Christensen:2018iqi}. }.  The LVC has recently placed bounds on the amplitude of this background based on the results of the O2 run~\cite{LIGOScientific:2019vic}. An upgraded aLIGO to full design sensitivity, together with Virgo, and KAGRA, may aid in detecting such a signal in the near future\footnote{For the duration of this paper, we refer to this system of detectors as $2^{nd}$ generation or HLVK.}.

While this is a good example of what can be gleamed from the sGWB, our work here will focus on testing GR with sGWB from stellar-mass binary black hole (BBH) sources which can be observed in the frequency band sensitive to ground based detectors ($\sim 30$ Hz).
Testing GR with the sGWB is only sensitive to the amplitude corrections\footnote{sGWB have also been used to probe additional GW polarizations~\cite{Abbott:2018utx}}. This is useful for, as an example, theories with parity violation that allows amplitude birefringence~\cite{Alexander:2007kv,Yunes:2010yf,Yagi:2017zhb}.

The possibility of testing GR with BBH sGWB has been proposed by Maselli, \textit{et. al.}~\cite{Maselli:2016ekw} within the ppE framework assuming that the non-GR correction is smaller than the GR contribution. The authors carried out a likelihood, model selection analysis to see what is the required non-GR amplitude correction such that the non-GR model is more preferred over the GR one. They mainly considered a phenomenological inspiral-merger-ringdown waveform~\cite{Ajith:2007kx}, though they also considered a possibility of using a simple power-law model for the sGWB spectrum. They used a network of second-generation (2G) detectors and focused on corrections entering at 0--1 post-Newtonian (PN) order relative to GR.
Their results indicated that indeed non-GR effects may be seen in the sGWB signal.  The work effectively assumed that the background GR spectrum was known \textit{a priori}, when comparing with the spectrum of theories beyond GR.

We extend this important work of~\cite{Maselli:2016ekw} in various ways. First, we carry out a parameter estimation study via Fisher method instead of a model selection analysis. We include not only the ppE parameter but also GR parameters (the amplitude and the slope of the spectrum). This means that we do not make an assumption that we know the GR spectrum \textit{a priori}, and automatically take into account correlations between GR and non-GR parameters. 
Second, we study a wider range on PN parameters than~\cite{Maselli:2016ekw} by considering PN orders ranging in between $\left(-4,4\right)$.
Third, we map our results to specific theories of modified gravity, namely non-commutative (NC) gravity~\cite{Harikumar_2006,Kobakhidze:2016cqh} and Varying-G (VG) theories~\cite{Tahura:2018zuq,Yunes:2009bv}.
Fourth, we consider not only a network of 2G detectors, but also that of third-generation (3G) detectors.
Lastly, we estimate systematic errors on the ppE parameter due to mismodeling of the sGWB spectrum with a simple power-law model.

For the purpose of this work, we will make some standard assumptions.  First, we assume the sGWB is isotropic (i.e. there is uniformity in the signal across the sky)~\cite{TheLIGOScientific:2016dpb}\footnote{While there is expected to be an anisotropic component based on the population of e.g. white-dwarf binaries within the Milky-Way galaxy~\cite{Adams:2013qma}, we will not consider these sources here as we restrict our analysis to ground-based detectors.}  Second, we assume that the sGWB is stationary.  That is, the information contained within the sGWB does not vary much during the time of observation.  Third, we expect the sGWB to be Gaussian in its distribution.  This was shown to be a favored model when the number of GW sources was high~\cite{Cornish:2015pda,Christensen:2018iqi,Allen:1996vm}. Lastly, we assume that the sGWB is unpolarized. That is, there is statistically equal amount of plus and cross modes in the signal\footnote{In non-GR theories, there can be additional polarizations that we do not consider in this paper. See e.g.~\cite{Nishizawa:2009bf,Isi:2018miq} for analyses of non-GR polarizations in sGWB.}.

\begin{figure}
	\includegraphics[width=\linewidth]{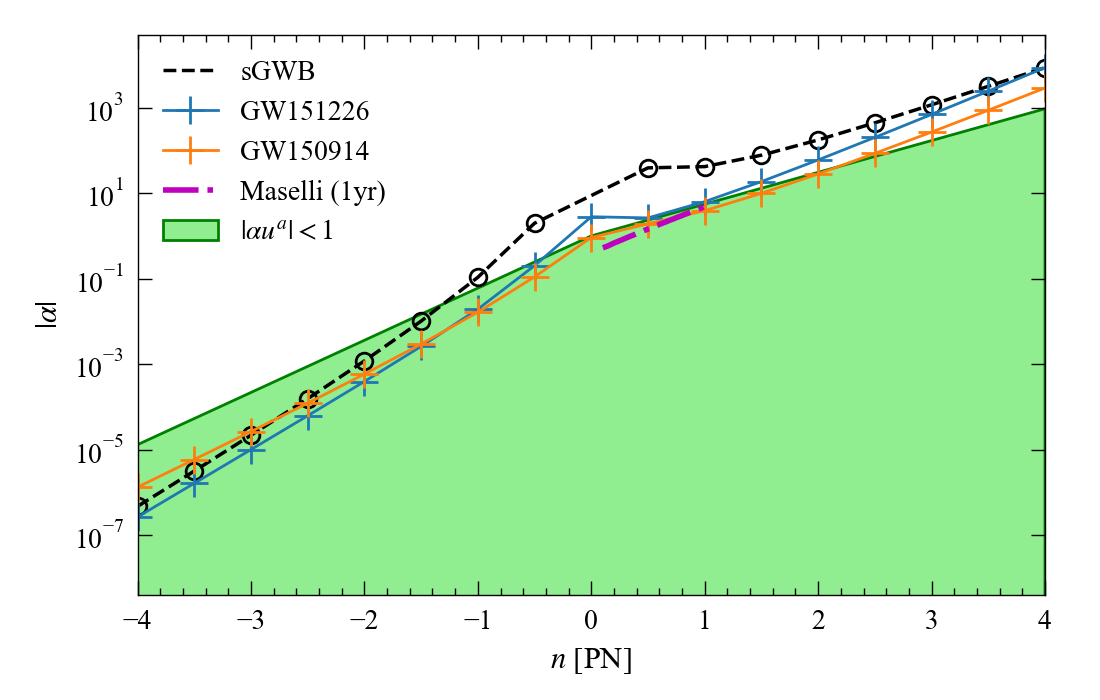}
    \caption{Upper bound on the amplitude ppE parameter $\alpha$ from sGWB (black circles) assuming a 1yr observation with a network of 2G detectors, compared with previous results from individual GW events~\cite{Tahura:2019dgr} (blue and orange crosses) and the previous work of Maselli \textit{et. al.}~\cite{Maselli:2016ekw} in magenta.  The green shaded region shows the valid region under small non-GR approximation imposed in the ppE formalism. We see that while placing limits on $\alpha$ is possible with the sGWB when the non-GR correction enters at $-1$PN order or lower, individual detections with higher SNR provide a better place for investigating theories beyond GR.    Notice we are unable to place bounds at 0 PN due to a degeneracy.}
    \label{fig:Results1}
\end{figure}

We present the main results of our analysis in Fig.~\ref{fig:Results1}.  We find that assuming a power law formalism for the sGWB can place bounds on the ppE amplitude parameter assuming the use of the HLVK ground based detectors when the corrections enter at a sufficiently negative PN order. However, bounds from existing individual events, such as GW150914 and GW151226, place stronger bounds, and the increased SNR from individual GW signals detected with future detectors will be able to place even more stringent bounds on the ppE parameter than can be seen with the sGWB alone.
Furthermore, we found that systematic errors on the ppE parameter due to the mismodeling of a power-law assumption is much smaller than statistical errors, giving a justfication of the model.  We also extend our analysis to 3G detectors, particularly the Einstein Telescope (ET)~\cite{Punturo:2010zz,Hild:2010id,Sathyaprakash:2012jk} assumed to be located at the current Virgo site, and Cosmic Explorer (CE)~\cite{Reitze:2019iox} located at the Hanford site.  However, the systematic errors of a power law are too large for this setup, and a better way to characterize the sGWB must be used for the ET-CE analysis.

The structure of the paper is as follows: Section~\ref{sec:sGWB_Intro} presents an introduction to the sGWB as well as the power law model we will use for our analysis.  Section~\ref{sec:PPE_Intro} will present the ppE formalism and how we extend its use to the sGWB.  Here, we also discuss the mapping to two specific modified theories of gravity.  In Sec.~\ref{sec:ParameterEstimation} we present our formulation for Fisher analysis and discuss how this is used to place bounds on the ppE parameters as well as how to find the systematic error incured from our power law assumption. Section~\ref{sec:Results} discusses the Fisher analysis results for our analysis.  Additionally, we apply these constraints to specific modified theories of gravity to predict the bounds on the ppE amplitude.  We conclude the paper in sec.~\ref{sec:Conclusions} with a discussion and possible future avenues for this work.
We will make use of the metric signature $(-,+,+,+)$ as presented in~\cite{Misner:1974qy} and the unit convention of $G=c=1$.

\section{GWB Spectral Energy Density in GR}
\label{sec:sGWB_Intro}
In this section, we discuss the basics of the sGWB.  We describe how we will calculate the energy density spectrum $\Omega_{\rm GWB}$, as well as introduce the Fisher methods we will make use of for predicting constraints we may place on the sGWB parameters.

\subsection{Fiducial Model}
We begin by defining the normalized sGWB spectral energy density as 
\begin{equation}
	\Omega_{\rm GWB} =\frac{1}{\rho_c}\frac{d \rho_{\rm GW}}{d \ln f}\,,
\end{equation}
where $\rho_c = 3H_0^2/(8\pi )$ is the critical density necessary to close the universe with $H_0$ the Hubble constant while $\rho_{\rm GW}$ is the energy density of GWs as a function of the GW frequency $f$.  We may show through the work of~\cite{Phinney:2001di} that $\Omega_{\rm GWB}$ from coalescing binaries may be rewritten as
\begin{equation}
\label{eq:PhinneyOmega}
	\Omega_{\rm GWB} = \frac{f}{H_0 \rho_c} \int_0^\infty \frac{R(z)}{(1+z)E(z)}\frac{dE_{\rm GW}[(1+z)f]}{df} dz\,,
\end{equation}
where $dE_{\rm GW}[(1+z)f]/df$ is the spectrum of GWs emitted by a source while accounting for the redshift $z$, $R(z)$ is the rate at which GWs are emitted per comoving volume, and $E(z)$ is a cosmological correction given by 
\begin{equation}
E(z) = \sqrt{\Omega_{\rm M}(1+z)^3+\Omega_{\Lambda}}
\end{equation}
for a flat universe.

\begin{figure}
	\includegraphics[width=\linewidth]{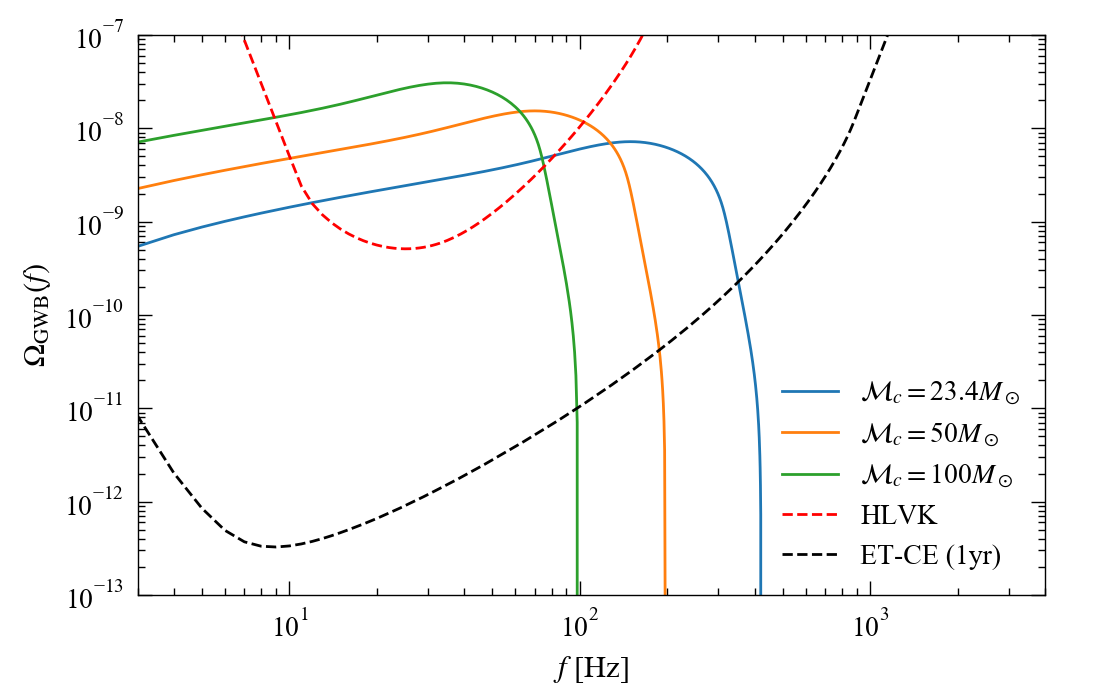}
    \caption{$\Omega_{\rm GWB}$ for various average chirp masses.  We include the power-law integrated sensitivity curve consisting of the Hanford, Livingston, Virgo, and KAGRA detectors. For simplicity, we assumed that these detectors all have aLIGO's design sensitivity which can be modeled analytically as shown in~\cite{Ajith:2011ec}. We also include our 3G setup consisting of the CE and ET detectors.  If the GW spectrum goes above the sensitivity curve in certain frequency range, the SNR is above unity.}
    \label{fig:EnergyBackgroundCurves}
\end{figure}

We consider a background consisting soley of BBH mergers. Taking the rate of these mergers to be $R(z)$, we apply the rate model discussed in~\cite{Callister:2016ewt}. In particular, we choose the BBH merger rate of the current universe as $R_0= 53.2_{-28.8}^{+58.5}$ Gpc$^{-3}$yr$^{-1}$~\cite{LIGOScientific:2018jsj}.  For the energy spectrum $dE_{\rm GW}[(1+z)f]/df$, we make use of the IMRPhenomB waveform\footnote{Bounds on $\alpha$ from GW150914 were compared in~\cite{Tahura:2019dgr} for the IMRPhenomB template and a more accurate IMRPhenomD template~\cite{Husa:2015iqa,Khan:2015jqa}. The two bounds agree when one uses only the inspiral signal, which is effectively what we consider in this paper.}~\cite{Ajith:2009bn}. 
Figure~\ref{fig:EnergyBackgroundCurves} shows $\Omega_{\rm GWB}$ as a function of frequency for various average chirp masses $\mathcal{M}_c = (m_1^3m_2^3/(m_1+m_2))^{1/5}$ with individual masses $m_1$ and $m_2$\footnote{The average chirp mass here refers to $\langle \mathcal M_c^{5/3} \rangle^{3/5}$~\cite{Callister:2016ewt}. For simplicity, we assumed all BBHs are equal-mass and non-spinning.}. The average chirp mass of $\mathcal M_c = 23.4M_\odot$ corresponds to the average chirp mass~\cite{Callister:2016ewt} for the BBHs detected in aLIGO's O1 and O2 runs~\cite{LIGOScientific:2018mvr}.  We have included the power-law integrated sensitivity curve for a network of 2G ground-based detectors given 15 months of observation time\footnote{This choice was made so that the SNR of the GW spectrum with fiducial values exceed the theshold value of 5, as we explain later.} calculated from~\cite{Thrane:2013oya}. We assume the network of detectors consists of the two aLIGO detectors at Livingston and Hanford, as well as the Virgo detector in Italy and the KAGRA detector in Japan.  
Our theoretical setup for the 3G configuration consists of the Einstein Telescope (ET)~\cite{Punturo:2010zz,Hild:2010id,Sathyaprakash:2012jk} located at the current Virgo site, and Cosmic Explorer (CE)~\cite{Reitze:2019iox} located at the Hanford site (see~\cite{Yagi:2017zhb} for more details).  The sensitivity curves for the ET and CE can be found on the LIGO Document Control Center~\cite{LIGODCCcitation}. For simplicity, we limit our observation time for the 3G setup to 1 year.

\subsection{Power Law Model}
Notice that from Fig.~\ref{fig:EnergyBackgroundCurves} the aspect of the sGWB spectrum is almost linear (on a log scale) through the power-law integrated sensitivity curve.  This line resembles a power law, which we may utilize to allow for easier computation, rather than focusing on the more detailed and computationally intensive phenomenological waveform.

\begin{figure}
    \centering
    \includegraphics[width=\linewidth]{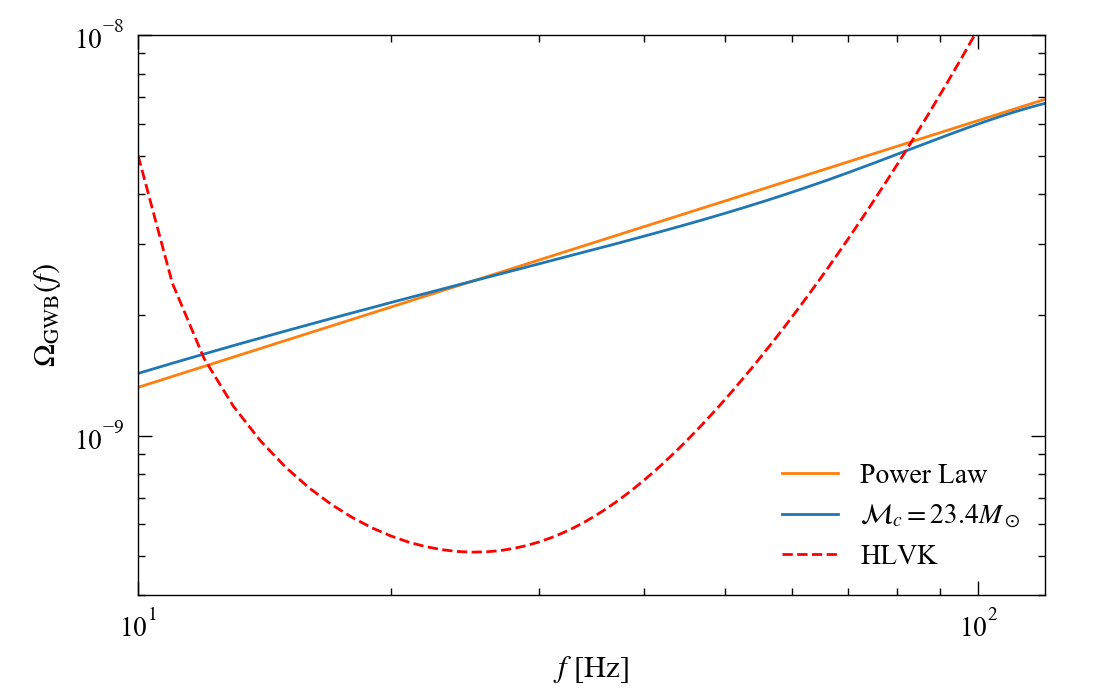}
    \caption{A comparison between the power law used in this paper and a phenomenological model for the GW spectrum. The former is constructed such that it has the same reference amplitude $\Omega_* = 2.43 \times 10^{-9}$ at $f_* = 25$Hz as the latter. Notice that in the sensitivity region for the 2G setup, there is good agreement in making this assumption.}
    \label{fig:PowerLawComparison}
\end{figure}

We denote such sGWB power law as~\cite{Kuroyanagi:2018csn}
\begin{equation}
	\label{eq:PowerLawIntroduction}
	\Omega_{\rm GWB} = \Omega_{*}\left( \frac{f}{f_{*}}\right)^{n_{*}}\,,
\end{equation}
where $\Omega_{*}$ is a reference amplitude at frequency $f_{*}$.  The parameter $n_{*}$ can be calculated by considering the Newtonian component of the binary inspiral phase, and is found to be $n_{*}=2/3$~\cite{Phinney:2001di}.  Recent detections have shown that for a reference frequency of 25 Hz, the energy density of the background is predicted as $\Omega_{\rm GWB} = 1.1^{+2.7}_{-0.9}\times 10^{-9}$~\cite{TheLIGOScientific:2016wyq}.  In Fig.~\ref{fig:PowerLawComparison}, we plot the power law and the phenomenological spectrum within the sensitivity curve of our 2G setup. Observe that the inspiral part of the latter can indeed be approximated by the former.

Making use of Eq.~\eqref{eq:PowerLawIntroduction} instead of Eq.~\eqref{eq:PhinneyOmega} will allow for easier parameterization when it comes to performing a Fisher analysis, as well as expanding into the ppE formalism.  For higher mass systems, the full waveform will need to be considered due to the merger frequency existing within our 2G setup sensitivity (see Fig.~\ref{fig:EnergyBackgroundCurves}).

\section{sGWB beyond GR}
\label{sec:PPE_Intro}
This section will go over a formalism for constructing the sGWB for non-GR theories in a theory-agnostic way.  We will begin with an overview of  the ppE formalism presented in~\cite{Yunes:2009ke}, followed by its use for the sGWB. We will also describe some example non-GR theories that can be mapped to the ppE framework.

\subsection{sGWB in ppE Formalism}
The ppE formalism was developed as a way to address the bias in GW astrophysics that GR is the correct theory of gravity.  Such biases could lead to inaccurate interpretations of observations by incorrectly associating the data with templates and waveforms which do not describe the correct physics (i.e. they assume GR is the ``correct" theory of gravity).  The ppE formalism was developed as a way to characterize various alternative theories of gravity and allow for model-independent tests of GR in an effort to overcome this inherent bias and allow for deviations of GR to be considered.

The ppE waveform in the frequency domain for a quasi-circular coalescing binary inspiral takes the form~\cite{Yunes:2009ke}
\[ \tilde{h}(f) = \tilde{h}^{\rm GR}(f) \left( 1 + \alpha u^a\right) e^{i \beta u^b}\,, \]
where $(\alpha, \beta, a, b)$ are the ppE parameters responsible for characterizing deviations away from GR, $u=(\pi \mathcal{M}_{c} f)^{1/3}$ is the effective relative velocity of black holes in a binary, and $a$ is related to the $n$th PN order by $n=a/2$. Notice that the above waveform reduces to the GR one $\tilde{h}^{\rm GR}$ in the limit $(\alpha,\beta) = (0,0)$.  As we will only concern ourselves with the inspiral portion of the binary in this paper, we direct the interested reader to~\cite{Yunes:2009ke} for more detailed explanations concerning the areas of the merger and ringdown.

Since we are making use of the inspiral portion of our waveform only, we may solve for $dE_{\rm GW}(f)/df$ in Eq.~\eqref{eq:PhinneyOmega} assuming a quasi-circular BH binary.  The energy flux is known in GR to be proportional to $|\tilde h(f)|^2$.  We use this, along with the ppE inspiral term being smaller than the GR one, to write the energy density as
\begin{equation}
\label{eq:PPE_Omega}
	\Omega_{\rm GWB} = \Omega_{\rm GWB}^{\rm GR} \left( 1 + 2 \alpha u^a\right) + \order{\alpha^2}\,.
\end{equation}
Notice that the ppE phase parameter $\beta$ does not enter in the above spectrum, and thus we are unable to place any constraints on such a parameter from sGWB observations.

\subsection{Mapping to Example non-GR Theories}
\label{sec:example}

One of the reasons why the ppE formalism can be useful is because the ppE waveform can be mapped to specific non-GR theories. See~\cite{Tahura:2018zuq} for examples of such a mapping, both for the amplitude and phase corrections. In this paper, we consider non-commutative (NC) gravity~\cite{Ardalan:1998ce,Seiberg:1999vs,Szabo:2001kg,Douglas:2001ba,Harikumar_2006,Kobakhidze:2016cqh} and varying-$G$ (VG) theories~\cite{Tahura:2018zuq,Yunes:2009bv} as representative theories. The leading ppE correction in the former enters at a positive PN order while that in the latter enters at a negative PN order. Below, we review these theories in turn, together with their respective expressions for $\alpha$ and $a$.

\subsubsection{Non-commutative gravity}

The first example is NC gravity. The theory has been proposed as a way to quantize spacetime coordinates ($\hat{x}_\alpha$) as proposed in~\cite{PhysRev.71.38}, and therefore it seeks to find agreement between quantum mechanics and GR. Such quantized spacetimes arise in the low energy field theory limit of string theory under certain backgrounds~\cite{Ardalan:1998ce,Seiberg:1999vs,Szabo:2001kg,Douglas:2001ba}.  These coordinate operators satisfy the canonical commutation relation
\begin{equation}
    \left[ \hat{x}_\mu , \hat{x}_\nu \right] = i \Theta_{\mu \nu}\,,
\end{equation}
where $\Theta_{\mu \nu}$ characterizes the ``fuzziness" in quantized spacetimes~\cite{Kobakhidze:2016cqh}.  We may introduce a new term which normalizes this in relation to Planck length ($l_p$) and time ($t_p$) such that~\cite{Kobakhidze:2016cqh}
\begin{equation}
    \Lambda^2 = \frac{\Theta^{0i}\Theta_{0i}}{(l_p t_p)^2}\,.
\end{equation}
With this normalized parameter, we may place bounds on NC gravity through the use of their relation to the ppE amplitude parameters provided in~\cite{Tahura:2018zuq} as
\begin{equation}
\label{eq:NC_bound_Eqnation}
    \alpha_{\rm NC} = -\frac{3}{8}\eta^{-4/5}\left( 2\eta-1\right) \Lambda^2\,, \quad a_{\rm NC} = 4\,,
\end{equation}
where $\eta = m_1m_2/(m_1+m_2)^2$ is the symmetric mass ratio of the binary. The correction enters at 2PN order in the waveform relative to GR. Kobakhidze et al.~\cite{Kobakhidze:2016cqh} derived corrections to the phase and derived a bound from GW150914 as $\sqrt{|\Lambda|} < 3.5$.

\subsubsection{Varying-$G$ Theories}

Theories in which a time varying gravitational constant come about often include those which violate the strong equivalence principle due to the presence of additional fields, like in scalar-tensor theories~\cite{Nordtvedt:1990zz,Yunes:2009bv}.  These VG theories also induce an anomalous acceleration in binary system~\cite{Tahura:2018zuq}.  Thus, given the parameters of the binary, we may be able to investigate the outcome of a time varying gravitational constant ($\dot{G}$).  In terms of a binary system, we find the ppE correction to VG theories to be~\cite{Tahura:2018zuq}\footnote{We assume that the time variation of $G$ that enters in Kepler's law is the same as that in GW luminosity.}
\begin{align}
    \alpha_{\rm VG} &= \frac{5}{512} \eta^{3/5} \dot{G}\left[ -7M + \left( s_1 + s_2\right)M \right. \nonumber \\
    & \left. \quad  + 13 \left(m_1s_1 + m_2s_2 \right) \right]\,,\\
    a _{\rm VG} &= -8\,,
\end{align}
where $M = m_1+m_2$ is the total mass of the system, $m_i$ are the component masses, and the sensitivities are defined to be $s_i =- (\dot{G}/m_i))(\delta m_i/\delta \dot G)$. 
The above correction enters at $-4$PN order in the waveform. Current bounds have been derived from Solar System experiments, binary pulsar and cosmological observations as $|\dot G| \lesssim 10^{-14} - 10^{-12} $/yr~\cite{Will:2014kxa,Copi:2003xd,Manchester:2015mda,Zhu:2015mdo,Zhu:2018etc,2018NatCo...9..289G}, while those from GW observations are much weaker~\cite{Yunes:2016jcc,Tahura:2019dgr}.

\section{Parameter Estimation}
\label{sec:ParameterEstimation}

We now explain how to perform a Fisher analysis on sGWB, following e.g.~\cite{Seto:2005qy,Kuroyanagi:2018csn}. 
The sGWB signal $s(t) = h(t) + n(t)$ consists of the GW strain $h$ and the noise $n$. The former is given by $h = h_{ij}F^{ij}$ where $F^{ij}$ is the beam-pattern function of a detector that depends on the sky position of the source and the polarization angle~\cite{Maggiore:1900zz}. $h_{ij}$ is further given  by
\begin{equation}
\label{eq:PlaneWaveExpansion}
	h_{ij}(t,\vec{x}) = \sum_{A} \int^\infty_{-\infty} df \int d^2\hat{n} \, \tilde{h}_{A}(f,\hat{n})\,e^A_{ij}(\hat{n})\, e^{-2\pi i f (t-\hat{n}\cdot \vec{x})}\,,
\end{equation}
where $\hat{n}$ is the direction from the detector to the GW source, A is the polarization ($+$ or $\times$ in GR), and $e^A_{ij}(\hat{n})$ is the polarization tensor for GWs.

\begin{figure}
	\includegraphics[width=\linewidth]{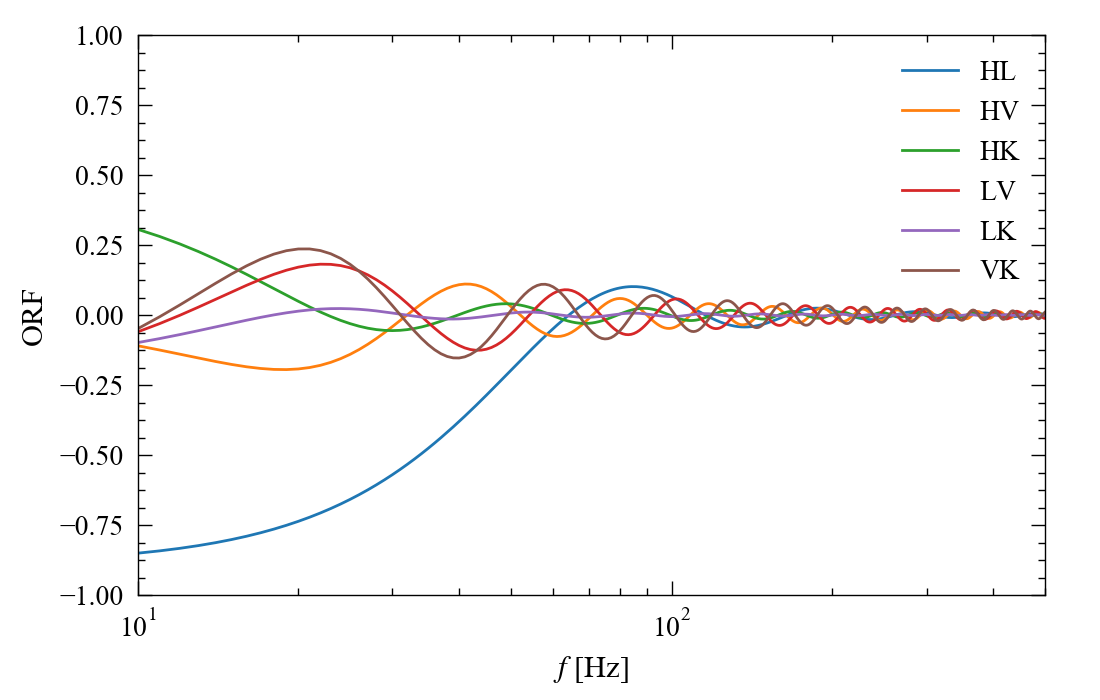}
	\includegraphics[width=\linewidth]{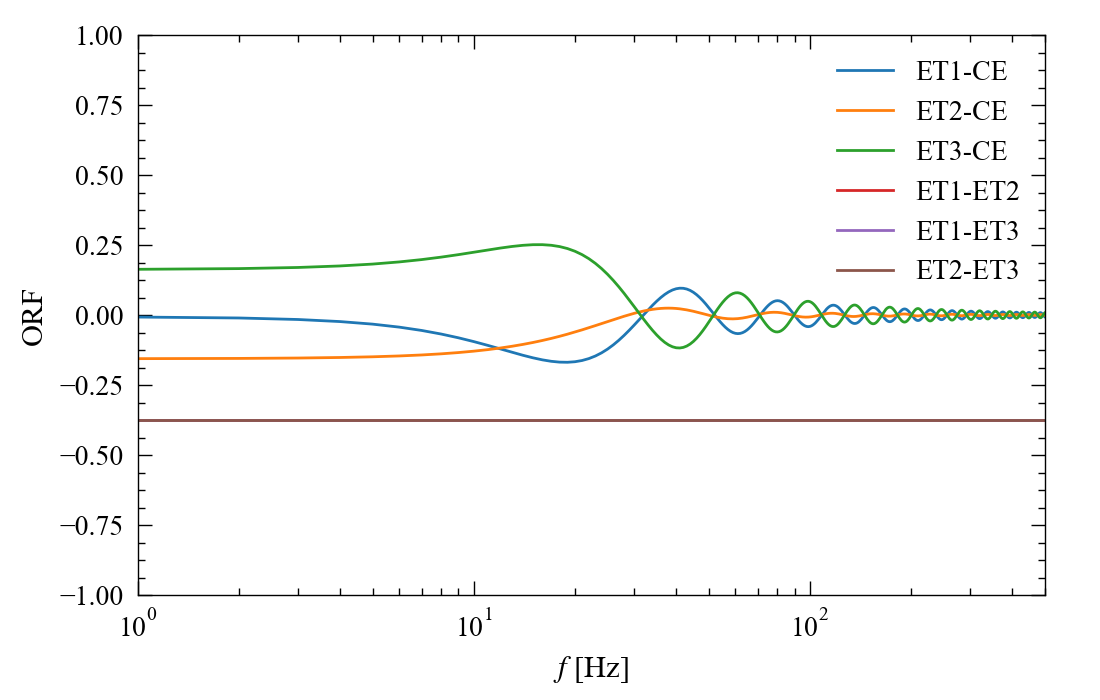}
    \caption{The overlap reduction function (ORF) for the various detectors in the 2G and 3G detector networks considered in this work.  Notice that the ORF for two independent interferometers in the ET detector is a constant.}
    \label{fig:ORF}
\end{figure}

The sGWB search is carried out by performing a cross-correlation between signals of detectors $I$ and $J$ using a filter function $Q(t,t')$ as
\begin{equation}
S = \int^{T_\mathrm{obs}/2}_{-T_\mathrm{obs}/2}  dt \int^{T_\mathrm{obs}/2}_{-T_\mathrm{obs}/2}  dt' s_I(t) s_J(t') Q(t,t')\,, 
\end{equation}
where $T_\mathrm{obs}$ is the observation time. We assume sGWB is isotropic, unpolarized, and stationary. Namely
\begin{align}
&\left\langle \tilde h_{P}^{*}(f, \hat{n}) \tilde h_{P'}(f^{\prime}, \hat{n}^{\prime})\right\rangle 
\nonumber \\
&=\frac{3 H_{0}^{2}}{32 \pi^{3}} \delta^{2}\left(\hat{n}-\hat{n}^{\prime}\right) \delta_{P P^{\prime}} \delta\left(f-f^{\prime}\right)|f|^{-3} \Omega_\mathrm{GWB}(|f|)\,,
\end{align}
where the angle brackets refer to an ensemble average. On the other hand, for stationary noise, the noise spectral density $S_n$ is defined as
\begin{equation}
\left\langle \tilde n^{*}(f) \tilde n\left(f^{\prime}\right)\right\rangle=\frac{1}{2} \delta\left(f-f^{\prime}\right) S_n(f)\,.
\end{equation}
Using these, the mean and variance of this correlated signal are given by
\begin{align}
\mu &=  \langle S \rangle \nonumber \\
&= \frac{3 H_{0}^{2}}{20 \pi^{2}} T_\mathrm{obs} \int_{-\infty}^{\infty} d f|f|^{-3} \gamma_{I J}(f) \Omega_{\mathrm{GWB}}(f) \tilde{Q}(f)\,, \\
\sigma^2 &= \langle S^2 \rangle - \langle S \rangle ^2 \nonumber \\
&= \frac{T_\mathrm{obs}}{4} \int_{-\infty}^{\infty} d f S_{n, I}(|f|) S_{n, J}(|f|)|\tilde{Q}(f)|^{2}\,,
\end{align}
where $S_n$ is the noise spectral density of a detector. $\gamma_{IJ}(f)$ is the overlap reduction function (ORF) between the $I$th and $J$th detectors that depend on detector's beam-pattern function. Figure~\ref{fig:ORF} presents the ORF for different combinations of 2G and 3G detectors.   $\tilde Q$ is the Fourier transform of $Q$. 

Let us next define the SNR and Fisher matrix.
The SNR, $\rho = \mu/\sigma$, is maximized when we choose the optimal filter function given by
\begin{equation}
\tilde Q(f) \propto \frac{\gamma_{I J}(|f|) \Omega_{\mathrm{GWB}}(|f|)}{|f|^{3} S_{n, I}(|f|) S_{n, J}(|f|)}\,.
\end{equation}
Then, the SNR is given by
\begin{equation}
\label{eq:SNR}
\rho = \frac{3 H_{0}^{2}}{10 \pi^{2}} \sqrt{2 T_\mathrm{obs}} \left[\sum_{I}^N \sum_{I<J}^N\int_{0}^{\infty} d f \frac{\left|\gamma_{I J}(f)\right|^{2} \Omega_{\mathrm{GWB}}(f)^{2}}{f^{6} S_{n, I}(f) S_{n, J}(f)}\right]^{1 / 2}\,,
\end{equation}
where $N$ is the number of detectors.
By maximizing the likelihood $\mathcal{L}$ of the correlated signal, one can estimate the statistical error on parameters $\theta^i$ as
\begin{equation}
\Delta^{(\mathrm{stat})} \theta^i = \sqrt{(\Gamma^{-1})_{ii}}\,,
\end{equation} 
where the Fisher matrix is defined by
\begin{align}
\label{eq:Fisher}
\Gamma_{ij} =& - \frac{d^2 \ln(\mathcal L)}{\partial \theta_i \partial \theta_j} \nonumber \\
= &2\left( \frac{3 H_0^2}{10 \pi^2}\right) T_{\rm obs} \nonumber \\
    &\times\sum_{I}^N \sum_{I<J}^N \int_0^\infty df \frac{|\gamma_{IJ}(f)|^2 \partial_i \Omega_{\rm GWB}(f) \partial_j \Omega_{\rm GWB}(f)}{f^6 S_{n,I}(f) S_{n,J}(f)}\,, \nonumber \\
\end{align}
with $\partial_i = \partial/\partial \theta^i$.
Practically, we change the integration range of Eqs.~\eqref{eq:SNR} and~\eqref{eq:Fisher} to $f_{\min}=10$ Hz and $f_{\max}=200$ Hz which we deem appropriate given these bounds are outside the integrated sensitivity curve seen in Fig.~\ref{fig:PowerLawComparison} for a network of 2G detectors.  For our 3G detector setup, we perform a similar analysis as with the 2G case with $f_{\min}=1$ Hz and $f_{\max}=f_m$, where $f_m$ is the merger frequency of the binary.  Note that any input outside of these bounds will not yield a signal due to the results being dominated by noise.

We will make use of Eq.~\eqref{eq:PowerLawIntroduction} for our Fisher analysis using Eq.~\eqref{eq:PPE_Omega}.  We assume a parameter space of 
\begin{equation}
\theta^i = \left( \ln\Omega_{*},  n_{*},\alpha\right)\,,
\end{equation}
 which accounts for both GR parameters $( \ln\Omega_{*},  n_{*})$ as well as a component beyond GR ($\alpha$).  This is in contrast to the previous work of~\cite{Maselli:2016ekw}, which effectively assumed the GR sGWB energy density was known \textit{a priori}. We use the fiducial values of $n_* = 2/3$ and $\alpha=0$, while we make use of a phenomenological waveform at $f_*=25$ Hz to find our fiducial value of $\Omega_*$ to be $2.43 \times 10^{-8}$. We assume the average chirp mass of $23.4M_\odot$, which corresponds to SNRs of 5.06 and 2530 for the network of 2G and 3G detectors respectively. 

In addition to the above Fisher analysis to find statistical errors on parameters, we perform a comparison of our method for finding the energy density to the results of the IMRPhenomB model to determine whether any systematic errors occur from our power law assumption for sGWB.  This is done to ensure the results of our initial assessment of the power law make sense.  Should these systematic errors outweigh the statistical errors, it shows our power law assumption is not valid for probing GR. Extending~\cite{Cutler:2007mi} that was originally developed for signals from isolated GW sources, systematic errors are defined as
\begin{align}
\label{eq:SystematicError}
     \Delta^\mathrm{(sys)} \theta^i =& (\Gamma_{ij})^{-1}\left[ 2\left( \frac{3 H_0^2}{10 \pi^2}\right) T_{\rm obs} \right. \nonumber \\
     & \times \left. \sum_{I}^N \sum_{I<J}^N\int_0^\infty df \frac{|\gamma_{IJ}(f)|^2 \Delta \Omega \partial_j \Omega_\mathrm{GWB}}{f^6 S_{n,I}(f) S_{n,J}(f)} \right]\,,
\end{align}
where $\Delta \Omega$ is the difference in sGWB energy density between two different models (in our case, the power-law and IMRPhenomB models).

\section{Results}
\label{sec:Results}

\begin{figure}
	\includegraphics[width=\linewidth]{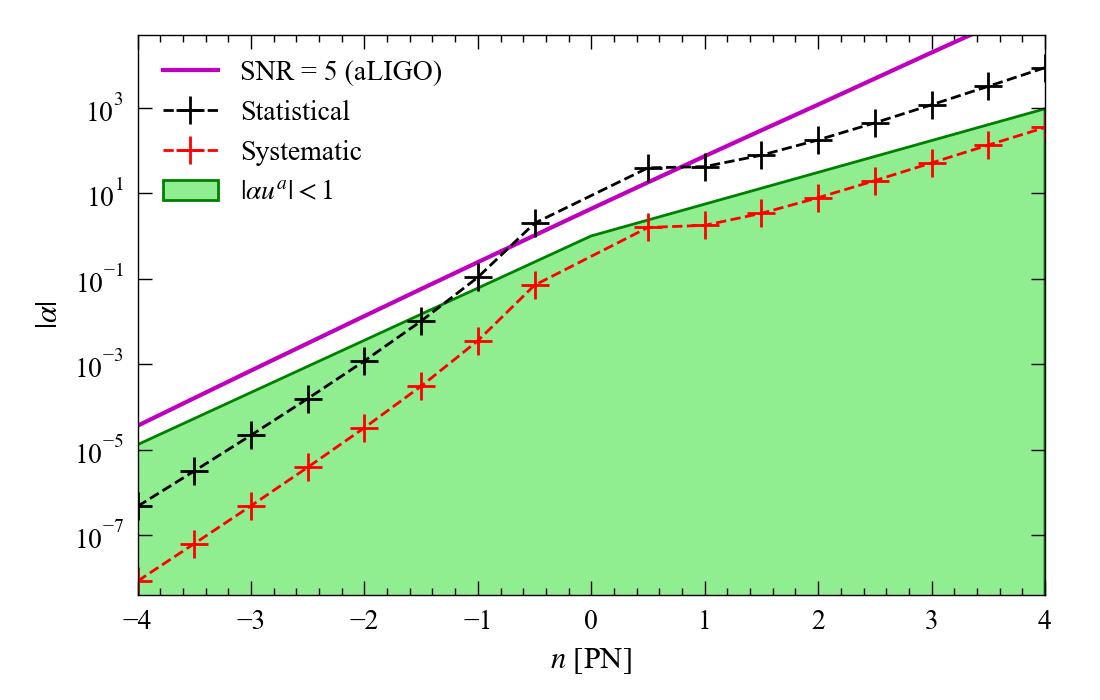}
    \caption{$\alpha$ bound obtained for the Fisher analysis (statistical) and the difference between the power law and IMRPhenomB waveform model (systematic). We choose the fiducial value of $\Omega_* = 2.43 \times 10^{-8}$ and assume an observation time of 15 months. Observe that the latter is smaller than the former, which justifies our use of the power-law model. We also show the value of $\alpha$ which may be found assuming an SNR threshold value of 5 for the sGWB detection using the realistic noise curves provided during the aLIGO's O1 and O2 run~\cite{LIGODCCcitationLivingston,LIGODCCcitationHanford,LIGODCCcitationVirgo}.}
    \label{fig:Results2}
\end{figure}

Let us now present our findings. We begin by investigating whether we can place bounds on $\alpha$ from the fact that sGWB was not detected during aLIGO's O1 and O2 run. When $\alpha$ is positive, the amplitude and SNR increase. Thus, one can estimate the value of $\alpha$ needed such that the SNR reaches the threshold value, which we choose to be 5. For this analysis, we consider aLIGO's noise spectral density  shown in~\cite{Abbott:2017oio}. We choose the observation time to be 13 months, which corresponds to the total duration of the O1 and O2 runs assuming the lasers were operational at all times.
Figure.~\ref{fig:Results2} shows a bound on $\alpha$ assuming a threshold SNR value of 5 for aLIGO's O1 and O2 runs.   This can be accomplished by making use of Eq.~\eqref{eq:SNR} and solving for the $\alpha$ value which yields an SNR of 5 for a given PN order.  The results of this calculation show that we are unable to place constraints on our ppE parameters (given the small ppE approximation) due to the fact that the $\alpha$ bound exists outside of our allowed regime.

The rest of this section focuses on presenting our findings on Fisher analyses with future detectors.
We first show the results for statistical errors on $\alpha$ with a network of 2G detectors with their sensitivities assumed to be identical and given analytically in~\cite{Ajith:2011ec}. 
Since we are assuming that the signal is consistent with GR (and thus the fiducial value of $\alpha$ is 0), such statistical errors correspond to projected bounds on $\alpha$ at each PN order once sGWB is detected.
Figure~\ref{fig:Results1} presents such upper bounds on $\alpha$ at each PN order from sGWB observations with a network of 2G detectors. 
Notice that we do not show bounds on the 0PN term due to a degeneracy between the energy density amplitude $\Omega_*$ and the ppE parameter $\alpha$ using the power law approximation.  
Since we are working within the assumption that the non-GR correction is smaller than the GR contribution, our bound is only valid in the green-shaded region which corresponds to $|\alpha u^a| < 1$ that is evaluated at the frequency for which our power law intersects the integrated sensitivity curve.  Since there are two points, we choose the frequency which provides more restrictive bounds on $\alpha$ (hence the discontinuity in the curve at 0 PN).  Figure~\ref{fig:Results1} shows that one can only place meaningful bounds when the non-GR correction enters at $n \leq -1$. Unfortunately, these bounds from sGWB are weaker than those already found through a similar Fisher analysis performed with GW150914~\cite{Tahura:2019dgr} and GW151226 in most cases. This suggests that isolated binaries can more efficiently probe non-GR effects in GW amplitudes.

For completeness, we map sGWB bounds to parameters in NC and VG gravity described in Sec.~\ref{sec:example} and compare them with existing bounds, including those from GW150914 and GW151226.
The results are summarized in Table~\ref{table:ParameterComparison}.  For simplicity, we have assumed the binary system in question is composed of two equal-mass black holes with a system chirp mass $23.4 M_{\odot}$ and sensitivities $s_1=s_2=0$.
\begin{table*}[]
\begin{tabular}{| c | c | c | c | c |}
\hline
 Parameter & sGWB (This Work) & GW150914 & GW151226 & Current Bound \\
 \hline
 $\sqrt{|\Lambda|}$ & 4.3  & 3.5~\cite{Kobakhidze:2016cqh}, 2.25\footnote{\label{TableNote} This bound was found through the phase correction to the waveform obtained in~\cite{LIGOScientific:2019fpa} by making use of a parameter estimation study where a single non-GR phase correction was allowed to vary, and calculated by making use of the fractional NC deviation from GR presented in~\cite{Kobakhidze:2016cqh}.  The bounds from~\cite{Kobakhidze:2016cqh} made use of a parameter estimation study where multiple non-GR parameters were allowed to vary, hence the slightly weaker bound.} & 1.96\footnoteref{TableNote} & 1.96--3.5~\cite{Kobakhidze:2016cqh} \\
 \hline
$|\dot{G}|$  [$10^{-12}$ /yr] & $2.127 \times 10^{18}$ & $5.4 \times 10^{18}$~\cite{Yunes:2016jcc} & $1.7 \times 10^{17}$~\cite{Yunes:2016jcc} & 0.04--1~\cite{Will:2014kxa,Copi:2003xd,Manchester:2015mda,Zhu:2015mdo,Zhu:2018etc,2018NatCo...9..289G}\\
 \hline
\end{tabular}
\caption{
The projected bounds on the parameters of two theories from future sGWB observations found in this work in comparison to previous results.
}
\label{table:ParameterComparison}
\end{table*}
We find that our results agree with what we expect to find from Fig.~\ref{fig:Results1}.  No sGWB bounds can be placed on these theories that are an improvement of the current best bound or the bound from individual GW sources like GW150914. We can thus say from this analysis that while our approach may be utilized to place limits on various modified theories of gravity, other approaches for testing GR may be more effective in placing stronger limits.

We next study the amount of systematic errors on $\alpha$ due to incomplete modeling of the power-law spectrum. We here assume that the true signal follows the GW spectrum obtained with IMRPhenomB waveform, and estimate systematic errors on $\alpha$ using the power-law model instead.
Using Eq.~\eqref{eq:SystematicError}, we find that the statistical error in determining the upper bound on $\alpha$ is greater than the systematic error (see. Fig.~\ref{fig:Results2}). This shows that systematic errors are under control and thus can be neglected.
Therefore, we are justified in making this assumption of a power law as the GW energy density background.

It may also be useful to look at a network of 3G detectors to determine whether any additional bounds may be placed on $\alpha$.  
We perform the same analysis as before, this time however with the ET-CE configuration of GW detectors.  
\begin{figure}
	\includegraphics[width=\linewidth]{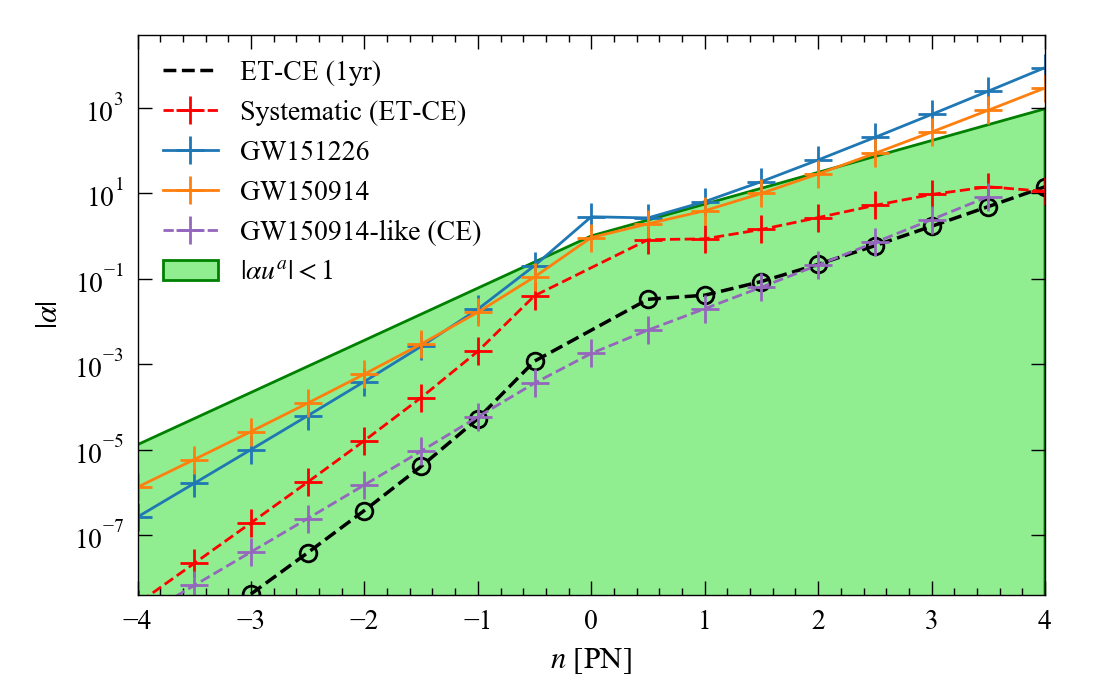}
    \caption{Similar to Fig.~\ref{fig:Results1} but for a network of 3G detectors. We have also included the CE bound found via a Fisher analysis for GW150914-like events presented in~\cite{Carson:2019kkh}.  Notice that unlike the 2G case, systematic error now dominates the error budget. 
    }
    \label{fig:3GResults1}
\end{figure}
The sensitivity curve of our 3G detectors is shown in Fig.~\ref{fig:EnergyBackgroundCurves}\footnote{The result for the power-integrated curve differs slightly from~\cite{Yagi:2017zhb} due to the fact the noise spectrum for ET in their work was made assuming the spectrum of CE, while we make use of more current ET models.}.    Notice that with the 3G detectors we will be sensitive to not only the inspiral, but to the merger and ringdown phases as well.  We again only consider the inspiral, and therefore terminate our integration at the merger frequency found from~\cite{Ajith:2009bn}.

Figure~\ref{fig:3GResults1} shows the results of our future bounds on $\alpha$ considering the inspiral phase of a BBH system with the 3G detectors observing for 1 year.  We find that our small perturbation limit of the ppE expansion is valid for all PN values and that stronger bounds can, in principle, be placed on the ppE amplitude than with current 2G detectors. We also see that the bounds on $\alpha$ from sGWB are now much stronger than those from GW150914 and GW151226. Of course, once the ET-CE system is turned on, we expect to find that individual sources are also louder and the bounds to become stronger (like the purple curve in Fig.~\ref{fig:3GResults1}). Thus we expect individual GW sources to be more useful for probing non-GR effects in the amplitude than with sGWB even for 3G detectors.

However, one needs to go beyond the power-law model for probing GR with sGWB using 3G detectors. This is because the systematic error for the sGWB is greater than the statistical error found via the Fisher analysis, as shown in Fig.~\ref{fig:3GResults1}.  The reason for this is that the statistical error is decreased due to the increased sensitivity of the detectors and a greater SNR of the signal.  On the other hand, the systematic error, that of making the power law approximation over using a phenomenological waveform model, very nearly remains constant as that used in the 2G setup.  This is because such systematic error is independent of the overall scaling of the detector sensitivity, and therefore we should not expect much change in the systematics between the 2G and 3G setups.

\section{Conclusions}
\label{sec:Conclusions}
We carried out a parameter estimation study involving the energy density of the sGWB assuming a power law spectrum modified by a ppE correction term.  Our results show that while we may indeed place limits on the ppE amplitude parameter by studying the sGWB, stronger bounds may be placed by making use of individual, higher SNR detections~\cite{Tahura:2019dgr}\footnote{We also note that we can access deviations in the phase when we use individual sources, which typically places stronger bounds on specific theories than from amplitude corrections~\cite{Tahura:2019dgr}.}.  In addition to this, we showed that, given the average chirp mass of the O1 and O2 observation runs of aLIGO and Virgo, the assumption of the background being made of inspiraling BBHs modeled as a power law will not significantly impact any study making use of 2G detectors, as the systematic error is less than statistical errors.

Making use of 3G detectors allow us to place stronger bounds on the ppE amplitude using a sGWB signal.  However, while the higher SNR from the ET-CE setup studied here leads to tighter constraints, the power law assumption leads to a systematic error that dominates over these ppE bounds, and therefore should not be used.

We may improve upon our results here by making use of a more robust GW spectrum model, such as the IMRPhenom waveform series~\cite{Ajith:2009bn,Husa:2015iqa,Khan:2015jqa,Pratten:2020fqn}, when dealing with the 3G detectors.  This will allow us to account for the merger and ringdown phases of the BBH events as well as lessen the systematic errors that were involved with our assumption of a power law model. It would be also interesting to study how sGWB from other sources, such as binary neutron star mergers~\cite{Abbott:2017xzg}, may affect the analysis here.  We may also carry out a similar analysis to the one shown here for other frequency regimes and other astrophysical sources of sGWB, such as those able to be detected by space-based detectors and pulsar timing arrays. Another direction for improvement is to carry out a Bayesian analysis, which is important given that the SNR for a network of 2G detectors considered here is only marginally detectable.

\acknowledgments
We would like to thank Sachiko Kuroyanagi for her helpful discussion and guidance with the Fisher analysis as well as Zack Carson for assistance with computational questions.
K.Y. would like to acknowledge support from NSF Award PHY-1806776, a Sloan Foundation Research Fellowship, the Ed Owens Fund, the COST Action GWverse CA16104 and JSPS KAKENHI Grants No. JP17H06358.

\bibliography{main.bib}

\end{document}